\documentstyle[aps,preprint,psfig]{revtex}
\begin{document}

\title{\bf Effect of Dissipation on Density Profile of One Dimensional Gas}

\author{Douglas A. Kurtze$^1$\thanks{E-mail address: 
   kurtze@golem.phys.ndsu.NoDak.edu}
 and Daniel C. Hong$^2$\thanks{E-mail address: 
   dh09@lehigh.edu} }

\address{$^1$ Department of Physics, University of Arizona,
   Tucson, Arizona 85721\\
and Department of Physics, North Dakota State University,
Fargo, North Dakota 58105\\
$^2$ Department of Physics, Lewis Laboratory, Lehigh University, 
   Bethlehem, Pennsylvania 18015}

\date{\today}
\maketitle

\begin{abstract}
We study the effect of dissipation on the density profile of a one-dimensonal 
gas that is subject to gravity.  The gas is in thermal equilibrium at
temperature T with a heat reservoir at the bottom wall.  Perturbative analysis 
of the Boltzmann equation reveals that the correction due to dissipation 
resulting from inelastic collisions is positive for $0 \le z < z_c$ and 
negative for $z > z_c$ with z the vertical coordinate.  The numerically
determined value for $z_c$ is $mgz_c/k_BT \approx 1.1613$, where $g$ is the 
gravitational constant and $m$ is the particle mass. 
\end{abstract}

\vskip 1.0 true cm
\section{Introduction}
\vskip 0.2 true cm

Granular materials are basically a collection of meso- to macroscopic particles 
that interact with each other via short range repulsive potentials [1].  For 
this reason, they may be considered as a very dense molecular gas, but with 
several key differences:  first, since the particles are macroscopic, gravity 
plays an important role and cannot be dismissed, as it can for a molecular gas. 
In fact, most of the unique features of granular materials disappear in the 
absence of gravity.  Second, collisions among the particles are inelastic, so
dissipation must be included in studies of the response of granular materials to
external stimuli.  It has recently been observed that dissipation appears to be 
responsible for the existence of a uniform cooling state and the associated 
clustering instability [2-7], and modifies the velocity distribution function 
from the usual Maxwellian to a power law or exponential function [8,9].  The
question of whether the appearance of the clustering instability is a generic 
phenomenon or an artifact of the hard sphere potential remains open; nonetheless
we note that the role of dissipation in nonequilibrium dynamics has always been
subtle, complex, and mathematically challenging.  

In this paper, we examine the effect of dissipation on the density profile of a 
one-dimensional granular gas within the framework of the Boltzmann equation.  In
so doing, we implicitly treat the grains as point particles, and thus ignore one
of the crucial aspects of granular materials, namely the excluded volume effect 
[10,11].  Neverthless, the outcome of this investigation is interesting and 
requires further study because it makes a nontrivial prediction: the dissipation
leads to an increase in density near the source and a decrease away from the 
source, with the dividing line appearing at a dimensionless height $g z / T$ of
approximately 1.1613.  This may have some relevance to the recent observation of
a clustering instability near the source in the presence of gravity [7].  
However, it remains to be seen whether our result persists in higher dimensions,
or survives the inclusion of the excluded volume effect.

\vskip 0.2 true cm
\section{Perturbative Analysis of the One-dimensional Boltzmann Equation}
\vskip 0.2 true cm

Consider a one-dimensional gas of $N$ identical classical point particles moving
along the $z$ axis above a wall at $z=0$, and acted upon by a constant 
gravitational field $g$ in the negative $z$ direction.  The particles collide 
with one another inelastically, with a coefficient of restitution $r < 1$.  Let
$Nf(u,z)\,du\,dz$ be the average number of particles found between heights $z$
and $z+dz$ with velocity between $u$ and $u+du$, in the steady state.  This
distribution function is given by the one-dimensional Boltzmann equation with 
dissipation [2],
 $$ \left( u {\partial \over \partial z} - g {\partial \over \partial u} \right)
     \, f(u,z) = C(u,z;q) - A(u,z).
  \eqno (1) $$
The collision terms on the right hand side are the ``creation'' term,
 $$ C(u,z;q) = \int\int |u'-u''| \, f(u',z) \, f(u'',z) \, 
                     \delta [u-qu'-(1-q)u''] \, du' \, du'',
  \eqno (2) $$
which gives the rate at which particles collide at $z$ and come out of the
collision with velocity $u$, and the ``annihilation'' term,
 $$ A(u,z) = f(u,z) \, \int |u'-u| \, f(u',z) \, du' = C(u,z;0),
  \eqno (3) $$
which is the rate at which particles with velocity $u$ collide at $z$ and
emerge with some other velocity.  The parameter $q$ is related to the
coefficient of restitution by $r = 1-2q$.  

We assume that the bottom wall at $z=0$ maintains the particles there in some
given velocity distribution $f_b(u)$.  This roughly means that the particle
distribution $f$ obeys the boundary condition $f(u,z=0) = f_b(u)$ at the bottom
wall, but we must be a bit more careful than this.  Since the particles are not
allowed into the region of negative $z$, the bottom wall can only propel 
particles upward.  Particles at $z=0+$ with positive velocities $u$ must have
been launched by the bottom wall, but those with negative velocities must have 
either undergone collisions or at least been in flight for a finite time since
they were last in contact with the bottom wall.  Therefore the velocity 
distribution of these latter particles is not $f_b(u)$, but instead is something
which was established by the dynamics described by the Boltzmann equation
itself.  Only the former particles, then, those with positive velocities, must
be in the velocity distribution imposed by the bottom wall.  Thus the correct
boundary condition is
 $$ f(u,0) = f_b(u) \qquad \hbox{for} \qquad u \ge 0.
  \eqno (4) $$
The other boundary conditions are the obvious ones, namely that $f$ should
vanish for $z \to \infty$ and for $|u| \to \infty$.

We now define a new parameter
 $$ \epsilon \equiv {q \over 1-q} = {1-r \over 1+r},
  \eqno (5) $$
which is small when the collisions are only slightly dissipative, i.e., when
the coefficient of restitution is only slightly less than unity.  Writing the
creation term (2) in terms of $\epsilon$ and carrying out the integral over 
$u''$ gives
 $$ C(u,z;q) = (1+\epsilon)^2 \int |u'-u| \, f(u',z) \, f(u+\epsilon[u-u'],z) 
                                 \, du'.
  \eqno (6) $$
We now expand this in powers of $\epsilon$ and substitute it into the Boltzmann 
equation (1) to get
 $$ \left( u {\partial \over \partial z} - g {\partial \over \partial u} \right)
     \, f(u,z) = \epsilon \int |u'-u| \, f(u',z) \, \left[ 2f(u,z) + 
        (u-u'){\partial f \over \partial u} \right] \, du' + $$
 $$  \epsilon^2 \int |u'-u| \, f(u',z) \, \left[ f(u,z) + 
        2(u-u'){\partial f \over \partial u} + 
        {1\over2}(u-u')^2{\partial^2 f \over \partial u^2} \right] \, du'
     + \cdots.
  \eqno (7) $$
This is the starting point of our calculation.

Let us write the solution of (7) as a power series,
 $$ f(u,z) = f_0(u,z) + \epsilon f_1(u,z) + \epsilon^2 f_2(u,z) + \cdots
  \eqno (8) $$
At zeroth order, $f_0$ must satisfy
 $$ u \, {\partial f_0 \over \partial z} - g \, {\partial f_0 \over \partial u} 
    = 0.
  \eqno (9) $$
This is satisfied identically by any function of the energy variable
 $$ E(u,z) = {1\over2}u^2 + gz.
  \eqno (10) $$
Thus to zeroth order we may satisfy the Boltmann equation and the boundary
condition (5) by writing
 $$ f_0(u,z) = f_b(\sqrt{2E(u,z)}).
  \eqno (11) $$
This is the complete solution to the problem in the case of elastic collisions.
Note that there would have been {\it no\/} solution to this problem had we 
attempted to impose the boundary condition (5) for {\it all\/} velocities $u$, 
with a function $f_b$ which is not even.

We will most often be interested in an assemblage which is in thermal
equilibrium at $z=0$.  In this case we have
 $$ f_0(u,z) = {\cal N} \exp(-E/T) = {\cal N} \exp(-(u^2+2gz)/2T),
  \eqno (12) $$
where the temperature is $m T / k_B$, where $m$ is the particle mass and $k_B$ 
is Boltmann's constant, and the normalization constant ${\cal N}$ is
 $$ {\cal N} = g / \sqrt{2 \pi T^3}.
  \eqno (13) $$

The zeroth order density profile is obtained by integrating over velocity,
 $$ \rho_0(z) = N \int_{\infty}^{\infty} f_0(u,z) du.
  \eqno (14) $$
In the thermal equilibrium case (12), this becomes
 $$ \rho_0(z) = (N g / T) \exp(-gz/T).
  \eqno (15) $$

We now calculate the first-order correction $f_1$ to the distribution
function.  From (7), we see that this must satisfy
 $$ \left( u {\partial \over \partial z} - g {\partial \over \partial u} \right)
     \, f_1(u,z) = \int |u'-u| \, f_0(u',z) \, \left[ 2f_0(u,z) + 
        (u-u'){\partial f_0 \over \partial u} \right] \, du'.
  \eqno (16) $$
The right side of this equation is the $u$-derivative of
$f_0(u,z) \int (u-u') |u-u'| f_0(u',z) \, du'$, but we have not found this
intriguing fact to be very useful for our purposes.  Note, however, that
$f_0(u)$ is an even function of $u$, and therefore the right side of (16) is
also even in $u$.  Since the operator on the left side is odd in $u$, this
implies that solving (16) will lead to a function which is odd in $u$.
Since we find the density profile by integrating over all $u$, this function
will not change the density from the zeroth order result (14).  However, the
operator in (16) also annihilates all functions of $E(u,z)$, so $f_1$ can
also contain any function of $E$.  Such a function is even in $u$, and so
does contribute to the density profile.

For the thermal equilibrium case, we can calculate $f_1$ explicitly.  First
note that the $z$ dependence of the right side of (16) consists entirely of
a factor $\exp(-2gz/T)$ coming from the two $f_0$ factors in the integrand.
To take advantage of this, we write $f_1$ in the form
 $$ f_1(u,z) = h_1(E(u,z)) - g^{-1} f_0^2(E(u,z)) F_1(u).
  \eqno (17) $$
Note that $F_1$ is independent of $z$.  Substituting this expression, and
the explicit form (12) of $f_0$, into (16) then gives
 $$ {dF_1 \over du} = \int_{-\infty}^\infty |u'-u|
     \left[ 2 + {u(u'-u) \over T} \right]
     \exp\left(-{u'^2-u^2 \over 2T}\right) \, du'.
  \eqno (18) $$
Changing integration variables from $u'$ to $v=u'-u$ and doing some algebra, 
we find that this can be written as
 $$ {dF_1 \over du} = 2 \int_0^\infty v \left( 4 - {v^2 \over T} \right)
     \exp\left(-{v^2 \over 2T}\right) \hbox{cosh}\left( {uv \over T} \right)
     \, dv,
  \eqno (19) $$
so we finally have
 $$ F_1(u) = 2T \int_0^\infty \left( 4 - {v^2 \over T} \right)
     \exp\left(-{v^2 \over 2T}\right) \hbox{sinh}\left( {uv \over T} \right)
     \, dv.
  \eqno (20) $$
This can also be written out explicitly in terms of error functions.

We must now determine the function $h_1(E)$, which forms the even part of $f_1$.
This is slightly more subtle than it first appears.  The first thing we should
do is apply the boundary condition (4) at $z=0$, thus maintaining the thermal
equilibrium distribution of particles coming upward from the bottom wall.
Thus we would choose $h_1(E)$ to be $g^{-1} f_0^2(E) F_1(\sqrt{2E})$.
However, as we will see below, this has a problem:  the resulting density
profile has fewer than $N$ particles.  In order to remedy this situation, we
must change the normalization of $f_0$, or equivalently include an extra
term proportional to $f_0(E)$ in $h_1(E)$, with a coefficient chosen to make
the total number of particles in the system again equal to $N$.  This amounts
to replacing the boundary condition (4) with a slightly more general condition
of the form
 $$ f(u,0) \propto f_b(u) \qquad \hbox{for} \qquad u \ge 0,
  \eqno (21) $$
with the proportionality constant chosen so that $f$ is properly normalized.
Thus we maintain the correct number of particles, and the {\it functional
form\/} of the distribution of particles launched from the bottom wall.  This
relaxation of condition (4) is physically reasonable, because the normalization
of the distribution $f_b$ is not directly observable.

We now have an expression for the distribution function $f$ correct to first
order in the dissipation parameter $\epsilon$,
 $$ f(u,z) = (1 + \epsilon {\cal N}_1) f_0(E) +
           \epsilon g^{-1} f_0^2(E) [F_1(\sqrt{2E}) - F_1(u)] + O(\epsilon^2),
  \eqno (22) $$
where $E$ is given by (10), $f_0$ by (12) and (13), $F_1$ by (20), and the
constant ${\cal N}_1$ is yet to be determined.  The density profile is
obtained by integrating over $u$; since $F_1(u)$ is an odd function it drops
out, leaving us with
 $$ N^{-1} \rho(z) = (1 + \epsilon {\cal N}_1) (g/T) \exp(-gz/T) +
           \epsilon g^{-1} \int f_0^2(E) F_1(\sqrt{2E}) \, du + O(\epsilon^2).
  \eqno (23) $$
Integrating this equation over all $z$ gives unity on the left side, so we see
that ${\cal N}_1$ must be given by
 $$ {\cal N}_1 = - g^{-1} \int_0^\infty dz \int_{-\infty}^\infty du 
                   f_0^2(E) F_1(\sqrt{2E}).
  \eqno (24) $$
Remarkably, the integrals on the right can be evaluated analytically, giving
the result
 $$ {\cal N}_1 = 2/\pi.
  \eqno (25) $$
Using this result in (23), we obtain the central result of this paper:  an
expression for the first-order correction to the density profile due to
dissipation.  After rescaling the integration variables $u$ and $v$ by factors
of $T^{1/2}$, we get
 $$ N^{-1} \rho_1(z) = {2 \over \pi} {g \over T} \left\{ \exp(-gz/T) \right. $$
$$\left. + \exp(-2gz/T) \int_0^\infty du \int_0^\infty dv (4-v^2)
    \exp(-u^2) \exp(-v^2/2) \hbox{sinh}(v\sqrt{u^2+2(gz/T)}) \right\}.
  \eqno (26) $$

We have evaluated (26) numerically as a function of $gz/T$ after first doing
the integral over $v$ analytically (in terms of error functions).  The result
is shown in Figure 1.  The most important feature of $\rho_1(z)$ is the fact
that it is {\it positive\/} near the source and {\it negative\/} away from the 
source, with a dividing line at $z_c = 1.1613 T/g$.  This is somewhat larger 
than $T/g$, which is the average $z$ coordinate of the particles when their
collisions are elastic.  Thus below $z_c$ the dissipation enhances the density 
profile, while above $z_c$ it suppresses it.  We might expect this behavior,
since the dissipation in the collisions removes mechanical energy from the 
system.  The effect of the thermal reservoir at the bottom wall is independent
of the characteristics of the collisions, so the particles leaving the bottom
wall are in the same distribution whether the collisions are elastic or not.
However, when their collisions are inelastic, every collision reduces the
total mechanical energy of the assemblage, thus reducing the average height
which the particles reach.  This may have some relevance to the recent 
experimental observation of the clustering instability by Kudrolli et al. [7], 
who noticed  the migration of the clustering instability along the direction of 
gravity and toward the driving source.
%

We note also that the perturbative correction to the density profile reaches
a minimum of $-0.206$ at $gz/T \approx 2.22$, and it also has a small maximum
of $1.0261$ at $gz/T \approx 0.037$.  We have not yet identified the physical 
mechanism responsible for the presence of this maximum.

It remains to be seen whether the enhancement of the density profile near the 
source in the presence of dissipation will persist in higher dimensions, or if
it will survive the inclusion of excluded volume effects in the model.
Future studies that elucidate these questions may provide interesting new
insights into the role of dissipation in granular dynamics.

\vfill\eject

\noindent {\bf References}

\noindent [1] H.M. Jaeger, S. R. Nagel, and R.P. Behringer, Rev. Mod. Phys. 
{\bf 68}, 1259 (1996).

\noindent [2] S. McNamara and W. R. Young, Phys. Fluids A {\bf 4}, 496 (1992); 
{\bf 5}, 34 (1993); S. McNamara, {\it ibid}. {\bf 5}, 3056 (1993).

\noindent [3] M. A. Hopkins and M. Y. Louge, Phys. Fluids A {\bf 3}, 4 (1990);
I. Goldirsch and G. Zanetti, Phys. Rev. Lett. {\bf 70}, 1619 (1993).

\noindent [4] B. Bernu and R. Mazighi, J. Phys. A: Math. Gen. {\bf 23}, 5745 
(1990).

\noindent [5] Y. Du, H. Li, and L.P. Kadanoff, Phys. Rev. Lett. {\bf 74}, 1268 
(1995).

\noindent [6] E. E. Esipov and T. Poschel, J. Stat. Phys. {\bf 86}, 1385 (1997).

\noindent [7] A. Kudrolli, M. Wolpert, and J. P. Gollub, Phys. Rev. Lett. 
{\bf 78}, 1383 (1997).

\noindent [8] J. Brey, F. Moreno, and J.W. Duffty, Phys. Rev. E. {\bf 54}, 445 
(1996).

\noindent [9] E. L. Grossman, T. Zhou, and E. Ben-Naim, preprint 
Cond-matt/9607165

\noindent [10] H. Hayakawa and D. C. Hong, Phys. Rev. Lett. {\bf 78}, 2764 
(1997).

\noindent [11] E. Clement and J. Rachjenbach, Europhys. Lett. {\bf 16}, 133 
(1991).

\vfill\eject

\noindent Figure Captions

Fig. 1.  The first-order correction $\rho_1(\tilde z)$ to the density profile, 
given by (26), plotted as a function of the dimensionless height $\tilde z = 
gz/T$.  The value at $\tilde z = 0$ is 1.  The graph then rises to a small
maximum at $\tilde z = 0.0372$ with a height of $1.0261$, crosses zero at
$\tilde z = 1.1613$, and reaches a minimum at $\tilde z = 2.2217$ with a value 
of $-0.2057$.  For large $\tilde z$ it approaches the axis as
$-2 \tilde z \exp(-\tilde z)$.

\vfill\eject

\vskip 1.0 true cm
\centerline{\hbox{
\psfig{figure=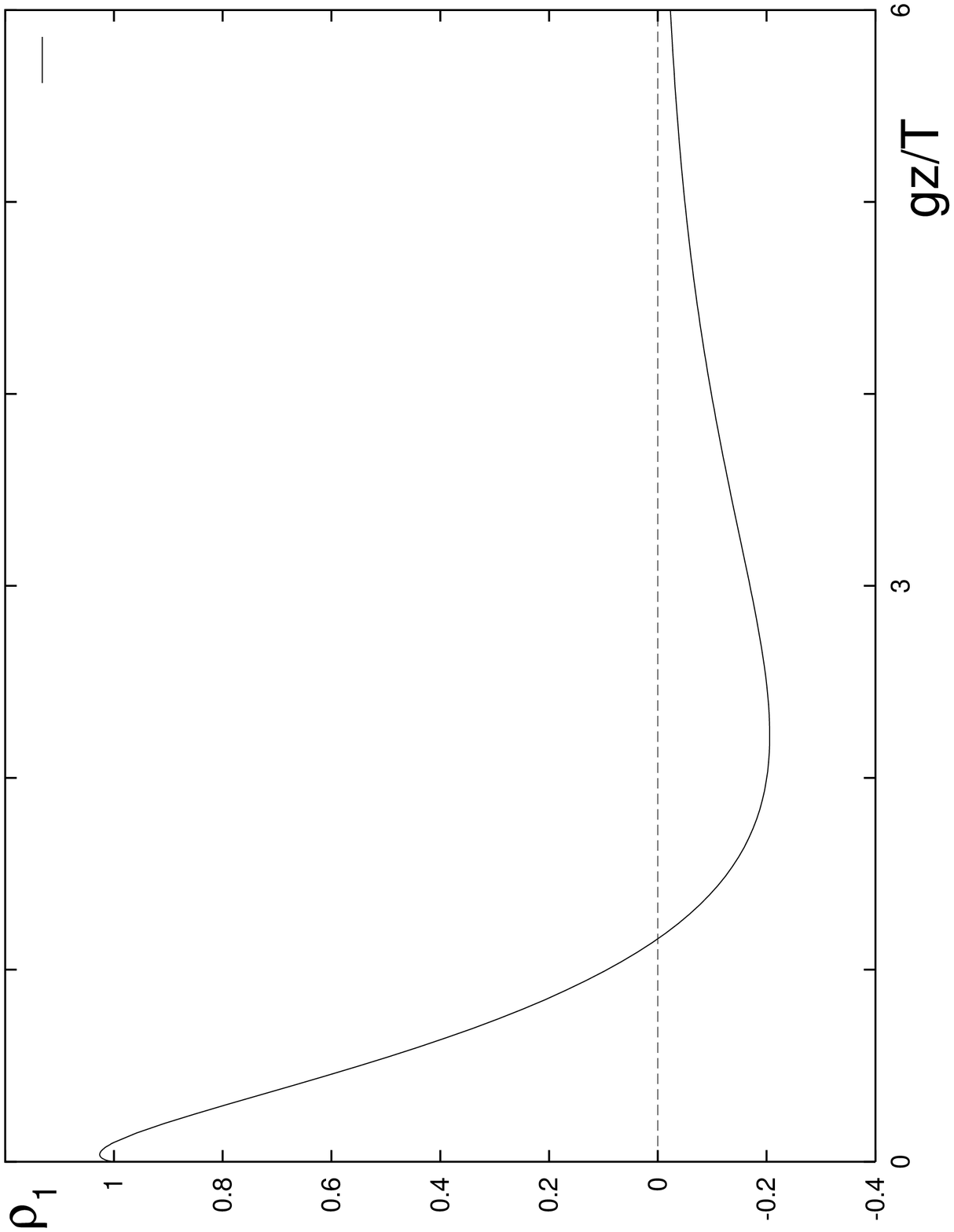}
}}

\end{document}